\documentclass{quantumview}
\usepackage[utf8]{inputenc}
\usepackage[english]{babel}
\usepackage[T1]{fontenc}
\usepackage{amsmath}
\usepackage{hyperref}
\usepackage{fixltx2e}
\usepackage[numbers,sort&compress]{natbib}
\usepackage{graphicx}
\usepackage{ragged2e}
\begin{document}

\title{Measurement-Device-Independent Quantum Secret Sharing}

\author{Xiao-Qiu~Cai}
\affiliation{School of Mathematical Science, Luoyang Normal University, Luoyang,China}
\author{Zi-Fan~Liu}
\affiliation{School of Mathematical Science, Luoyang Normal University, Luoyang,China}
\affiliation{School of Mathematics and Information Science, Henan Normal University, Xinxiang,China}
\author{Tian-Yin~Wang}
\affiliation{School of Mathematical Science, Luoyang Normal University, Luoyang,China}
\affiliation{School of Mathematics and Information Science, Henan Normal University, Xinxiang,China}
\orcid{0000-0003-1533-8015}

\maketitle
Quantum secret sharing plays an important role in quantum communications and secure multiparty computation. In this paper, we present a new measurement-device-independent quantum secret sharing protocol, which can double the space distance between the dealer and each sharer for quantum transmission compared with prior works. Furthermore, it is experimentally feasible with current technology for requiring just three-particle Greenberger-Horne-Zeilinger states and Bell state measurements.
\section{Introduction}
Quantum secret sharing (QSS) is a basic primitive in quantum cryptography, which allows a secret to be shared among participants in such a way that only the authorized sets can reconstruct it while no unauthorized set can acquire information on the secret even if they have unlimited computing resources\cite{Hillery_1999}. Due to this speciality, QSS plays an important role in secure key management, sensitive data protection and secure multiparty computation etc~\cite{9076282,9233436}. Consequently, many proposals for QSS~\cite{ Long_2002,PhysRevLett.95.230505,Qin_2007,Yaoyao2018Quantum,2018A,PhysRevA.103.032410,2022Theory} have been reported in both theoretical and experimental aspects since Hillery et al proposed the first QSS protocol based on Greenberger-Horne-Zeilinger (GHZ) states in 1999~\cite{Hillery_1999}.
Although the unconditional security of QSS has been strictly proved in theory, there still exists security leaks in implementation, whereby many cryptographic attacks were presented by taking advantage of the imperfections of practical devices. The technique of measurement-device-independent (MDI) is an effective way to deal with the security leaks existing in practical measurement apparatus. In 2015, the first measurement-device-independent quantum secret sharing (MDI-QSS) protocol was proposed by Fu et al~\cite{Fu_2015}. Subsequently, many novel proposals for MDI-QSS were presented~\cite{2016,Yaoyao2018Quantum,2019Measurement}. In 2020, a deterministic MDI-QSS protocol was reported~\cite{2020Deterministic}, which can avoid corresponding waste, but it is vulnerable to a special participant attack~\cite{2021SCPMA..6460321Y,2021A}. Recently, Ju et al reported a novel MDI-QSS protocol with hyper-encoding~\cite{Ju_2022}.

Inspired by the ideas in~\cite{PhysRevLett.95.200502,10.1063/1.4817672,CAI2022128226}, we give a new three-party MDI-QSS protocol, in which no sharer can recover the dealer's secret alone except that they cooperate with each other. Compared with the prior work, it is also detector-loophole-free and can prevent participant attack. Furthermore, the requirements for quantum states and their measurements are not rigorous, which make it experimentally feasible with current technology. Most importantly, it can double the space distance between the dealer and each sharer for quantum transmission by the way of introducing untrusted relays to perform Bell measurement.

\section{The MDI-QSS protocol}
Assume that the dealer Alice wants to distribute a secret $s$ including $n$ bits to two sharers Bob and Charlie while she requires that they recover the secret $s$ if and only if they cooperate with each other. The protocol includes the following steps (Fig. 1).

\begin{figure}

    \includegraphics[width=\textwidth]{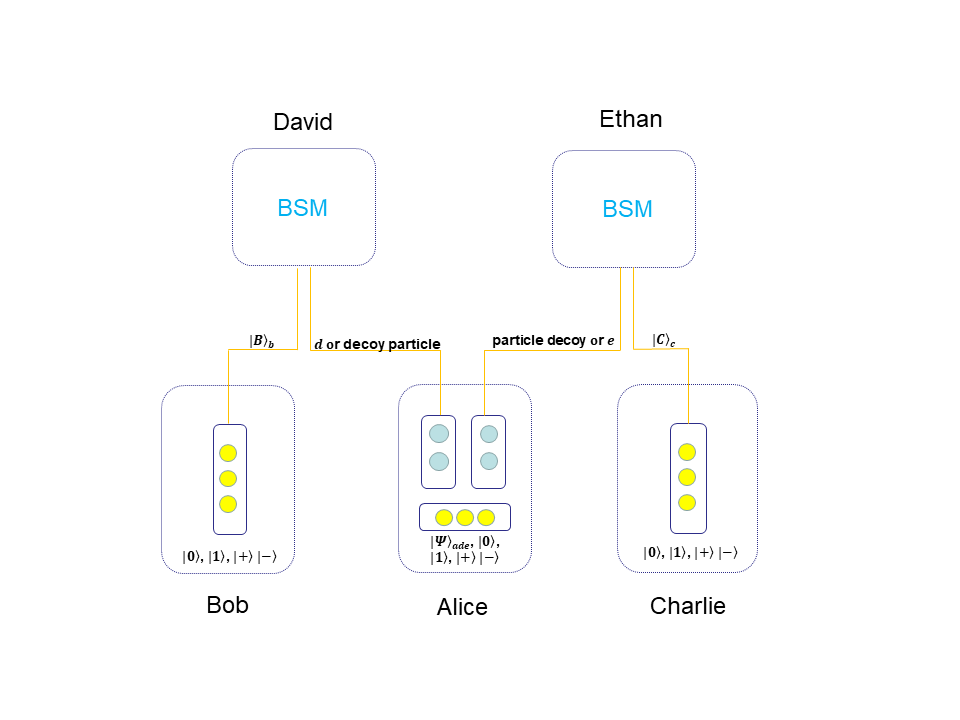}\\

  \caption{\textbf{The drawing of quantum part.}  The blue circles represent the $GHZ$ state, and the yellow circles represent the quantum states $|0\rangle$,$|1\rangle$,$|+\rangle$,$|-\rangle$. Bob and Charlie randomly chose  particles from $S$, and Alice randomly choses particles from  $GHZ$ or $S$. David and Ethan perform BSM on $|B\rangle_{b}, d$ and $|C\rangle_{c}, e$ and announce the measurement results afterwards. }\label{fig1}
\end{figure}

 \begin{itemize}
\item[(1)] Alice prepares a three-particle GHZ state
\begin{equation}\label{eqexpmuts}
|\Psi\rangle_{ade}=\frac{1}{\sqrt{2}}(|000\rangle+|111\rangle)_{ade}
\end{equation}
with the probability $p$, or two decoy states with the probability $1-p$, which are randomly chosen from a set $S=\{|0\rangle,|1\rangle,|+\rangle,|-\rangle\}$, here $|\pm\rangle=\frac{1}{\sqrt{2}}(|0\rangle\pm|1\rangle)$.

\item[(2)] In Step (1), when Alice prepares $|\Psi\rangle_{ade}$, she completes the next actions.

\item[(2.1)] She sends $d$ particle to David and $e$ particle to Ethan while keeping $a$ particle. At the same time, Bob and Charlie randomly chose $|B\rangle_{b}$ and $|C\rangle_{c}$ particles from $S$ and send them to David and Ethan, respectively.

\item[(2.2)] Once receiving the particles from Alice and Bob, David immediately performs a Bell state measurement (BSM) on them and broadcasts the outcome to Alice, Bob and Charlie, where the BSM is $\{|\varphi^{+}\rangle,|\varphi^{-}\rangle,|\psi^{+}\rangle,\\|\psi^{-}\rangle\}$, here $|\varphi^{\pm}\rangle=\frac{1}{\sqrt{2}}(|00\rangle\pm|11\rangle)$ and $|\psi^{\pm}\rangle=\frac{1}{\sqrt{2}}(|01\rangle\pm|10\rangle)$ denote four Bell states. Analogously, Ethan performs a BSM on the particles from Alice and Charlie and broadcasts the outcome.

\item[(2.3)] Bob and Charlie send their bases (Z-basis=$\{|0\rangle,|1\rangle\}$ or X-basis=$\{|+\rangle,|-\rangle\}$) to Alice.

\item[(2.4)] Once receiving the bases sent by sharers, if they are different, Alice does nothing; otherwise, she measures $a$ particle with the same basis.

\item[(3)] In Step (1), when Alice prepares decoy states, she completes the next actions.

\item[(3.1)] She sends two decoy states to David and Ethan, respectively. Two sharers perform the same actions as that in Step (2.1).

\item[(3.2)] Two relays also do the same actions as that in Step (2.2).

\item[(3.3)] Two sharers respectively send their state information not just the bases to Alice.

\item[(3.4)] When receiving Bob's information, if the bases of the decoy state and Bob's state are different, Alice takes no actions; otherwise, she checks whether there is an error in the quantum channels between her and David, or between Bob and David by the decoy state, Bob's state and the BSM outcome from David. For instance, when Bob's state is $|0\rangle$, the decoy state is $|1\rangle$, and David's BSM is $|\psi^{+}\rangle$ or $|\psi^{-}\rangle$, Alice finds no error; otherwise, she checks an error. Analogously, she judges whether there is an error in the quantum channel between her and Ethan, or between Charlie and Ethan.

\item[(4)] Steps (1)-(3) are repeated $m$ rounds.

\item[(5)] Alice estimates the error rate of quantum channels by the records made in Step (3.4), and if it is more than the preset threshold, she restarts the protocol from Step(1); Otherwise, she uses the measurement outcomes on all the $a$ particles with X-basis while both sharers' states are also in X-basis to generate a raw key $k_{r}$, and the measurement outcomes on the other $a$ particles with Z-basis are used for quantifying the amount of privacy amplification while both sharers' states are also in Z-basis. Noted that the measurement outcomes $|0\rangle,|+\rangle$ denote classical bit 0, and $|1\rangle,|-\rangle$ represent classical bit 1.

\item[(6)] Alice performs error correcting and privacy amplification on the raw key $k_{r}$ with two sharers in the similar way in~\cite{PhysRevLett.95.200502}, and then distill a final key $k$ including $n$ bits.

\item[(7)] Alice encrypts the secret $s$ with the key $k$, i.e., $c=k\oplus s$, and then sends the ciphertext $c$ to one sharer (e.g., Bob), where $\oplus$ denotes the addition of module 2.

\item[(8)] Two sharers cooperate to recover the key $k$ with their respective shares, and then obtain the secret $s$ by computing $s=k\oplus c$.
 \end{itemize}
\section{The analysis}

\subsection{Correctness}

\textbf{Theorem 1}. Suppose that all three participants are honest in the above MDI-QSS protocol, two sharers can recover the dealer's secret $s$ if they cooperate with each other.

Proof: Firstly, we show that if all participants perform the MDI-QSS protocol faithfully, Bob and Charlie can share a raw key $k_{r}$, which is generated by the dealer Alice. When the states prepared by both Bob and Charlie are in the X-basis, their relations with $|\Psi\rangle_{ade}$ will satisfy Tab. 1. From Tab. 1, we can find that if David and Ethan perform BSM on the $bd$ and $ce$ particles, respectively, then the $a$ particle will collapse to the state $|+\rangle$ or $|-\rangle$ with equal probability. Moreover, even if Bob or Charlie knows the BSM outcomes on the $bd$ and $ce$ particles, none of them can deduce the state of $a$ particle except that they exchange their states' information. For example, if Bob's state is $|+\rangle_{b}$ and the BSM outcomes of the $bd$ and $ce$ particles are $|\varphi^{+}\rangle|\varphi^{+}\rangle$, Bob cannot deduce the state of the $a$ particle except that he knows Charlie's state $|+\rangle_{c}$ or $|-\rangle_{c}$. Consequently, Bob and Charlie can cooperate to share a raw key $k_{r}$ with Alice.

Secondly, the final key $k$ is generated by distilling the raw key $k_{r}$, and the postprocessing of both error correcting and privacy amplification does not influence the correctness, which can be directly derived from Ref.~\cite{PhysRevLett.95.200502}. Therefore, Bob and Charlie can recover Alice's secret $s$ if they cooperate with each other.

To sum up, two sharers can recover the dealer's secret $s$ if they cooperate with each other if all three participants are honest.

\begin{table}
\caption{Correlations among the measurement outcomes with X-basis}
\begin{center}

  \renewcommand\arraystretch{1}
  \setlength{\tabcolsep}{1mm}
  \begin{tabular}{|c|c|c|}
    \hline
    Bob/Charlie  & Davlid/Ethan  & Alice   \\ \hline
    $|+\rangle|+\rangle$ & $|\varphi^{+}\rangle|\varphi^{+}\rangle,|\varphi^{+}\rangle|\psi^{+}\rangle,|\varphi^{-}\rangle|\varphi^{-}\rangle,|\varphi^{-}\rangle|\psi^{-}\rangle$ & $|+\rangle$ \\
                         & $|\psi^{+}\rangle|\varphi^{+}\rangle,|\psi^{+}\rangle|\psi^{+}\rangle,|\psi^{-}\rangle|\varphi^{-}\rangle,|\psi^{-}\rangle|\psi^{-}\rangle$             & \\ \hline
    $|+\rangle|+\rangle$ & $|\varphi^{+}\rangle|\varphi^{-}\rangle,|\varphi^{+}\rangle|\psi^{-}\rangle,|\varphi^{-}\rangle|\varphi^{+}\rangle,|\varphi^{-}\rangle|\psi^{+}\rangle$ & $|-\rangle$ \\
                         & $|\psi^{+}\rangle|\varphi^{-}\rangle,|\psi^{+}\rangle|\psi^{-}\rangle,|\psi^{-}\rangle|\varphi^{+}\rangle,|\psi^{-}\rangle|\psi^{+}\rangle$             & \\ \hline
    $|+\rangle|-\rangle$ & $|\varphi^{+}\rangle|\varphi^{-}\rangle,|\varphi^{+}\rangle|\psi^{-}\rangle,|\varphi^{-}\rangle|\varphi^{+}\rangle,|\varphi^{-}\rangle|\psi^{+}\rangle$ & $|+\rangle$ \\
                         & $|\psi^{+}\rangle|\varphi^{-}\rangle,|\psi^{+}\rangle|\psi^{-}\rangle,|\psi^{-}\rangle|\varphi^{+}\rangle,|\psi^{-}\rangle|\psi^{+}\rangle$             & \\ \hline
    $|+\rangle|-\rangle$ & $|\varphi^{+}\rangle|\varphi^{+}\rangle,|\varphi^{+}\rangle|\psi^{+}\rangle,|\varphi^{-}\rangle|\varphi^{-}\rangle,|\varphi^{-}\rangle|\psi^{-}\rangle$ & $|-\rangle$ \\
                         & $|\psi^{+}\rangle|\varphi^{+}\rangle,|\psi^{+}\rangle|\psi^{+}\rangle,|\psi^{-}\rangle|\varphi^{-}\rangle,|\psi^{-}\rangle|\psi^{-}\rangle$              & \\ \hline
    $|-\rangle|+\rangle$ & $|\varphi^{+}\rangle|\varphi^{-}\rangle,|\varphi^{+}\rangle|\psi^{-}\rangle,|\varphi^{-}\rangle|\varphi^{+}\rangle,|\varphi^{-}\rangle|\psi^{+}\rangle$ & $|+\rangle$ \\
                         & $|\psi^{+}\rangle|\varphi^{-}\rangle,|\psi^{+}\rangle|\psi^{-}\rangle,|\psi^{-}\rangle|\varphi^{+}\rangle,|\psi^{-}\rangle|\psi^{+}\rangle$             & \\ \hline
    $|-\rangle|+\rangle$ & $|\varphi^{+}\rangle|\varphi^{+}\rangle,|\varphi^{+}\rangle|\psi^{+}\rangle,|\varphi^{-}\rangle|\varphi^{-}\rangle,|\varphi^{-}\rangle|\psi^{-}\rangle$ & $|-\rangle$ \\
                         & $|\psi^{+}\rangle|\varphi^{+}\rangle,|\psi^{+}\rangle|\psi^{+}\rangle,|\psi^{-}\rangle|\varphi^{-}\rangle,|\psi^{-}\rangle|\psi^{-}\rangle$              & \\ \hline
    $|-\rangle|-\rangle$ & $|\varphi^{+}\rangle|\varphi^{+}\rangle,|\varphi^{+}\rangle|\psi^{+}\rangle,|\varphi^{-}\rangle|\varphi^{-}\rangle,|\varphi^{-}\rangle|\psi^{-}\rangle$ & $|+\rangle$ \\
                         & $|\psi^{+}\rangle|\varphi^{+}\rangle,|\psi^{+}\rangle|\psi^{+}\rangle,|\psi^{-}\rangle|\varphi^{-}\rangle,|\psi^{-}\rangle|\psi^{-}\rangle$             & \\ \hline
    $|-\rangle|-\rangle$ & $|\varphi^{+}\rangle|\varphi^{-}\rangle,|\varphi^{+}\rangle|\psi^{-}\rangle,|\varphi^{-}\rangle|\varphi^{+}\rangle,|\varphi^{-}\rangle|\psi^{+}\rangle$ & $|-\rangle$ \\
                         & $|\psi^{+}\rangle|\varphi^{-}\rangle,|\psi^{+}\rangle|\psi^{-}\rangle,|\psi^{-}\rangle|\varphi^{+}\rangle,|\psi^{-}\rangle|\psi^{+}\rangle$             & \\ \hline

  \end{tabular}
\end{center}
\end{table}

\subsection{Security}

As we know, if a cryptographic protocol is secure against participant attacks, then it must be also secure for external opponents~\cite{7962191,9491034}. This is also the case for QSS, and hence the main security goal of this protocol is to prevent the dishonest participants from deceiving.

\textbf{Theorem 2}. No unauthorized set can gain access to the dealer's secret $s$ even if they collude with each other in the above MDI-QSS protocol.

Proof: As shown in the MDI-QSS protocol, there is just one authorized set \{Bob, Charlie\}, which can recover the dealer's secret $s$ according to Theorem 1. Obviously, two participant sets \{Bob, David, Ethan\} and \{Charlie, David, Ethan\} have the most advantages. Accordingly, we only need to prove the impossibility for the two unauthorized sets to obtain the secret $s$ although there are lots of unauthorized sets.

Firstly, we show that Bob, David and Ethan cannot gain access to the dealer's secret $s$ even if they collude with each other. Obviously, the final key $k$ is necessary to recover the dealer's secret $s$, and therefore Bob, David and Ethan must get it. For the final key $k$ is generated by the postprocessing of the raw key $k_{r}$, they cannot gain access to $k$ in this process, which has been proven in~\cite{PhysRevLett.95.200502}. Consequently, Bob, David and Ethan must try to steal the raw key $k_{r}$. Nevertheless, Bob, David and Ethan can deduce no information on Alice's measurement outcomes by the resources (the BSM outcomes and Bob's state information) held by them from Table 1, which means that they cannot obtain the raw key $k_{r}$ through this way.

To obtain the raw key $k_{r}$, Bob, David and Ethan have to deduce the measurement outcomes on all the $a$ particles with
X-basis. They may have three ways. The first way is to get Charlie's state information and then deduce Alice's
measurement outcomes on all the $a$ particles with X-basis according to the correlations shown in Table 1. The second
way is to measure all the $d$ and $e$ particles with X-basis and then directly get Alice's measurement outcomes on all the
corresponding $a$ particles with X-basis by the correlations shown in Eq. (2).
\begin{eqnarray}
|\Psi\rangle_{ade}&=&\frac{1}{\sqrt{2}}(|000\rangle+|111\rangle)_{ade}\nonumber\\
                  &=&\frac{1}{2}[|+\rangle_{a}(|++\rangle_{de}+|--\rangle_{de})\nonumber\\
                  &+&|-\rangle_{a}(|+-\rangle_{de}+|-+\rangle_{de})]
\end{eqnarray}
The third way is to entangle an ancillary particle on each GHZ state  and then measure them to infer Alice's measurement outcomes on the $a$ particles with X-basis. Nevertheless, three ways are impossible for them. To get Charlie's state information, they can directly measure the $c$ particle with X-basis after receiving it in each round, then prepare a fake
one according to the measurement outcome and perform the protocol normally. They also can keep all the $c$ particles in their store and then perform a joint measurement on them to get information on Charlie's state until the protocol is completed. Nevertheless, in any case their deception cannot escape the security check of channels. In each round, the $c$ particle is randomly chosen from the set $S$, there is no way to discriminate it
according to quantum indistinguishable principle. Therefore, the probability that the fake particle replaced by them is
not the same as the original one is 50\%, in this case it will be detected with the probability 50\% if Alice performs the
security check, which means that in each round that Alice chooses the decoy particle and performs the security check,
the probability that David and Ethan can escape the security check of quantum channel is $50\%\times50\%=25\%$. Moreover, the number of rounds that Alice performs the security check is $(p-1)m$. By computing, the probability that they may escape the security check is
\begin{equation}\label{eqexpmuts1}
p_{s}=(25\%)^{(p-1)m}=\frac{1}{4^{(p-1)m}},
\end{equation}
which is exponentially close to 0 with the increase of $m$ and therefore is negligible in an ideal case. Another method to get
Charlie's state information is entangling an ancillary particle on each $c$ particle, and then measure it to exact some information on Charlie's states, which will face on the same difficulty and hence is not feasible. Therefore, the first way to deduce Alice's measurement outcomes on all the $a$ particles with X-basis is impossible by getting Charlie's state information. It seems that the second way is easy for Bob, David and Ethan because all the $d$ and $e$ particles are available, but the premise is that they must escape Alice's security check, which is also impossible because of the same difficulty. Finally, we analyze the impossibility of the third way. Without loss of generality, assume that the ancillary particle is $|E\rangle=\alpha|0\rangle+\beta|1\rangle$ and the general operation $U$ is defined as follows
\begin{eqnarray}
U|0\rangle|E\rangle&=&a|00\rangle+b|01\rangle+c|10\rangle+d|11\rangle, \\
U|1\rangle|E\rangle&=&a'|00\rangle+b'|01\rangle+c'|10\rangle+d'|11\rangle
\end{eqnarray}
where all the coefficients are complex numbers and satisfy the conditions $||\alpha||^{2}+||\beta||^{2}=1$, $||a||^{2}+||b||^{2}+||c||^{2}+||d||^{2}=1$ and $||a'||^{2}+||b'||^{2}+||c'||^{2}+||d'||^{2}=1$, here $||\cdot||$ denotes the modulus of a complex number. If Alice chooses to send a decoy state (e.g. $|0\rangle$) to Ethan for security check, Charlie's state (e.g., $|0\rangle$), the decoy state and the ancillary particle $|E\rangle$ will evolve in the state
\begin{eqnarray}
|0\rangle(U|0\rangle|E\rangle)&=|0\rangle(a|00\rangle+b|01\rangle+c|10\rangle+d|11\rangle)\nonumber\\
                             &=a|000\rangle+b|001\rangle+c|010\rangle+d|011\rangle.\nonumber\\
\end{eqnarray}
In order to avoid introducing error, the coefficients should satisfy $c=d=0$, which implies that $U|0\rangle|E\rangle=|0\rangle|E\rangle$. By similar
analysis, we can get $a'=b'=0$ and $U|1\rangle|E\rangle=|1\rangle|E\rangle$. Accordingly, the operation $U$ has no action on $|E\rangle$ and they can get no information on the $a$ particle by measuring the ancillary particle $|E\rangle$.

Secondly, we can prove that Bob, David and Ethan also cannot gain access to the dealer's secret $s$ even if they collude with each other by the similar analysis.

Based on the above analysis, we can conclude that only the authorized set \{Bob, Charlie\} can recover the dealer's secret $s$, but any unauthorized set cannot do it even if they collude with each other in the MDI-QSS protocol.
\section{Conclusion}

We give a new three-party MDI-QSS protocol and analyze its security, which is secure against all attacks in the detection system and participant attacks. Compared with the prior works, this protocol can double the space distance between the dealer and each sharer for quantum transmission. Furthermore, it is based on three-particle GHZ states and single-particle states while only BSM and single-particle state measurements are required, which make it experimentally feasible with current technology. We hope this work shed some light on the development of quantum cryptography.
\section{Acknowledgment}

We thank Prof. K. Chen very much for helpful discussion and suggestion. This work was supported by the National Natural Science Foundation of China (Grant Nos. 62272208, 62172196, 61902166), and the Natural Science Foundation of Henan Province, China (Grant No. 212300410062).

\bibliographystyle{quantum}
\bibliography{quantumview-template}

\end{document}